\documentclass[conference,letterpaper]{IEEEtran}

\usepackage[utf8]{inputenc} 
\usepackage[T1]{fontenc}
\usepackage{url}
\usepackage{ifthen}
\usepackage{cite}
\usepackage[cmex10]{amsmath} 

\usepackage{bm}
\usepackage{enumerate}
\usepackage{framed}
\usepackage{amssymb}
\usepackage{scalerel}
\usepackage{graphicx}

\newcommand{\mr}{\mathrm}
\newcommand{\BS}{\begin{subequations}}
\newcommand{\ES}{\end{subequations}}

\newtheorem{proposition}{Proposition}
\newtheorem{assumption}{Assumption}
\newtheorem{definition}{Definition}

\newtheorem{lemma}{Lemma}
 
\DeclareRobustCommand{\rchi}{{\mathpalette\irchi\relax}}
\newcommand{\irchi}[2]{\raisebox{\depth}{$#1\chi$}}

\begin{document}
\title{Overflow-Avoiding Memory AMP} 



\author{\IEEEauthorblockN{Shunqi~Huang\textsuperscript{\dag}, Lei~Liu\textsuperscript{\ddag},  and  Brian~M.~Kurkoski\textsuperscript{\dag}\\
  \textsuperscript{\dag}School of Information Science, Japan Institute of Science and Technology (JAIST), Nomi, Ishikawa, Japan \\
  \textsuperscript{\ddag}College of Information Science and Electronic Engineering, Zhejiang University, Hangzhou, Zhejiang, China \\
  Emails: \{shunqi.huang, kurkoski\}@jaist.ac.jp\textsuperscript{\dag}, lei\_liu@zju.edu.cn\textsuperscript{\ddag}
  }
}


\maketitle

\begin{abstract}
    Approximate Message Passing (AMP) type algorithms are widely used for signal recovery in high-dimensional noisy linear systems. Recently, a principle called Memory AMP (MAMP) was proposed. Leveraging this principle, the gradient descent MAMP (GD-MAMP) algorithm was designed, inheriting the strengths of AMP and OAMP/VAMP. In this paper, we first provide an overflow-avoiding GD-MAMP (OA-GD-MAMP) to address the overflow problem that arises from some intermediate variables exceeding the range of floating point numbers. Second, we develop a complexity-reduced GD-MAMP (CR-GD-MAMP) to reduce the number of matrix-vector products per iteration by $1/3$ (from $3$ to $2$) with little to no impact on the convergence speed.
\end{abstract}

\section{Introduction}
In this paper\footnote{This work was supported in part by the National Natural Science Foundation of China (NSFC) under Grants 62394292 and 62301485, in part by the Excellent Young Scientists Program (Overseas) of NSFC, and in part by the Research Fund of ZTE Corporation.}, we focus on the signal recovery problem from a noisy linear system $\bm{y} = \bm{Ax} + \bm{n}$, where $\bm{x} \in \mathbb{C}^{N}$ is the signal to be recovered, $\bm{y} \in \mathbb{C}^{M}$ is the known observation, $\bm{A} \in \mathbb{C}^{M \times N}$ is known, and $\bm{n}\!\sim\!\mathcal{CN}(\bm{0},\sigma^2\bm{I}_M)$ is Gaussian noise. In general, unless $\bm{x}$ is Gaussian, the optimal signal recovery is NP-hard \cite{verdu1985optimum, micciancio2001hardness}.

\subsection{Background}\label{Sec:background}
Approximate message passing (AMP) \cite{donoho2009message} was proposed for solving the signal recovery problem from noisy linear systems. AMP offers several key advantages. First, it boasts low complexity, requiring only $\mathcal{O}(MN)$ computations per iteration. Additionally, the mean square error (MSE) performance of AMP can be effectively tracked through the state evolution (SE) \cite{donoho2009message, bayati2011dynamics}. Furthermore, if $\bm{A}$ is IID Gaussian and the SE has a unique fixed point, AMP achieves minimum MSE (MMSE) optimality (Bayes optimality) for uncoded systems \cite{reeves2019replica, barbier2020mutual} and achieves capacity optimality for coded systems \cite{liu2021capacity}.

Nevertheless, AMP performs poorly or even diverges when confronted with matrices containing correlated entries \cite{manoel2014sparse, vila2015adaptive, rangan2017inference}. To overcome this limitation, several algorithms were developed. A variant of AMP based on a unitary transformation, called UTAMP \cite{guo2015approximate}, was proposed for correlated matrices. Though UTAMP exhibits promising performance, its Bayes optimality, convergence and SE have not been proven. Additionally, UTAMP requires singular value decomposition (SVD), leading to a high complexity $\mathcal{O}(M^2N)$. Another significant development is orthogonal/vector AMP (OAMP/VAMP) \cite{ma2017orthogonal, rangan2019vector}, which was proposed for right-unitarily-invariant matrices. The SE of OAMP/VAMP was conjectured in \cite{ma2017orthogonal}, and \cite{rangan2019vector, takeuchi2020rigorous} proved that this SE can track the performance of OAMP/VAMP. Crucially, when the SE has a unique fixed point, it was proved that OAMP/VAMP is replica Bayes-optimal for uncoded systems \cite{ma2017orthogonal, rangan2019vector, takeuchi2020rigorous} and replica capacity optimal for coded systems \cite{liu2021capacity, liu2021capacity_oamp}. However, the complexity of OAMP/VAMP is as high as $\mathcal{O}(M^3+M^2N)$ since the linear MMSE (LMMSE) estimator requires a matrix inversion. In a recent theoretical work \cite{liu2023oamp}, a generic iterative process (GIP) was introduced, highlighting the significance of the error orthogonality principle. AMP, OAMP/VAMP can be unified under the GIP framework \cite{liu2023oamp, takeuchi2019unified}. In addition, there is significant theoretical work in studying general AMP algorithms for unitarily-invariant matrices \cite{fan2022approximate}, as well as studying the universality class of sensing matrices \cite{dudeja2022spectral}.

To overcome the limitation of high complexity, a promising alternative called convolutional AMP (CAMP) \cite{takeuchi2021bayes} was proposed for right-unitarily-invariant matrices. CAMP offers a complexity of $\mathcal{O}(MN)$ per iteration, which is comparable to that of AMP. It was proved that CAMP is replica Bayes-optimal when its SE converges to a unique fixed point. However, CAMP may diverge for matrices with middle-to-large condition numbers. To solve the convergence problem of CAMP, a general framework for designing memory AMP-type algorithms, called memory AMP (MAMP), was proposed in \cite{liu2022memory}. MAMP unifies AMP, OAMP/VAMP, and CAMP under a single framework. Furthermore, a Bayes-optimal MAMP algorithm was designed with the utilization of a long-memory MF estimator. In this paper, the Bayes-optimal MAMP in \cite{liu2022memory} is called gradient descent MAMP (GD-MAMP) since its linear estimator can be viewed as a memory GD process combined with orthogonalization. The complexity of GD-MAMP is $\mathcal{O}(MN)$ per iteration. Most importantly, the convergence of SE for GD-MAMP is guaranteed by analytically optimized damping, and GD-MAMP achieves replica Bayes-optimality for uncoded systems when the SE has a unique fixed point. Another notable development is warm-started conjugate gradient VAMP (WS-CG-VAMP) \cite{skuratovs2022compressed}, which converges faster than GD-MAMP for matrices with large condition numbers, while maintaining comparable complexity. However, due to limited precision in practical systems, incorrect computations for orthogonalization parameters may cause WS-CG-VAMP to diverge. Although an approximation method that can avoid this problem was given in \cite{skuratovs2022compressed}, it is not effective for matrices with large condition numbers.

\subsection{Motivation and Contribution}
In GD-MAMP, some of the intermediate variables in the SE and orthogonalization processes of GD-MAMP exhibit exponential growth in the number of iterations, which may exceed the range of floating point numbers. This overflow problem of GD-MAMP causes the algorithm to not operate properly. In our previous work \cite{liu2022memory}, we avoided this problem through programming techniques and did not give a solution with the theoretical guarantee. It remains a question of how to solve the
overflow problem fundamentally. 

In addition, GD-MAMP in \cite{liu2022memory} requires four matrix-vector products, which is double that required in AMP. This fact leads to two interesting questions: Is it feasible to reduce the complexity of GD-MAMP? Can we develop a variant of GD-MAMP that only requires two matrix-vector products per iteration while maintaining comparable convergence speed?

In this paper, we answer these questions. The main contributions include:
\begin{itemize}
    \item 
    We develop an overflow-avoiding GD-MAMP (OA-GD-MAMP) with the known eigenvalues of $\bm{A}\bm{A}^{\rm H}$, which is equivalent to the original GD-MAMP while solving its overflow problem. When the eigenvalues of $\bm{A}\bm{A}^{\rm H}$ are unknown, we develop an alternative form of OA-GD-MAMP that demonstrates nearly identical convergence.
    \item 
    We simplify the original GD-MAMP, reducing the number of matrix-vector products required per iteration from four to three. More importantly, we develop a complexity-reduced GD-MAMP (CR-GD-MAMP), which only requires two matrix-vector products per iteration, while maintaining comparable convergence speed.
\end{itemize}


\section{Overview of MAMP and GD-MAMP}
In this section, we begin by introducing the problem formulation and outlining underlying assumptions. Next, we review the definition and properties of MAMP, along with an overview of the GD-MAMP algorithm.

\subsection{Assumptions}
The assumptions are listed as follows:
\begin{assumption} \label{Assum:1}
    The entries of $\bm{x}$ are IID with zero mean. The variance of $\bm{x}$ is normalized, i.e., $\tfrac{1}{N}{\rm E}\{\|\bm{x}\|^2\}=1$. For some $i > 0$, the $(2\!+\!i)$-th moments of $\bm{x}$ are finite.
\end{assumption}
\begin{assumption} \label{Assum:2}
    The system is large-scale, i.e., $M, N \to \infty$ with a fixed $\delta=M/N$.
\end{assumption}
\begin{assumption} \label{Assum:3}
    $\bm{A}$ is a right-unitarily-invariant matrix, meaning that for the singular value decomposition (SVD) $\bm{A} = \bm{U\Sigma}\bm{V}^{\rm H}$, $\bm{V}$ is Haar distributed and independent of $\bm{U\Sigma}$.
\end{assumption}
\begin{assumption} \label{Assum:4}
Let $\lambda_{\max}$ and $\lambda_{\min}$ be the maximum and minimum eigenvalues of $\bm{A}\bm{A}^{\rm H}$ respectively, and let $\lambda_t \equiv \tfrac{1}{N}{\rm tr}\{(\bm{A}\bm{A}^{\rm H})^t\}$. We assume that $\lambda_{\max}$, $\lambda_{\min}$, and $\lambda_t$ for $t < 2T$ are all known, where $T$ is the maximum number of GD-MAMP iterations. Furthermore, $\lim_{N\to\infty} \tfrac{1}{N}{\rm tr}\{\bm{A}\bm{A}^{\rm H}\} \overset{\rm a.s.}{=} 1$.
\end{assumption}

In the case that $\{\lambda_{\rm max}, \lambda_{\rm min}, \lambda_t\}$ are unavailable, a method to estimate them was given in \cite{liu2022memory}.

\subsection{Memory AMP}
\begin{definition}[Memory AMP \cite{liu2022memory}]\label{def:MAMP}
A memory AMP is an iterative process that consists of a memory linear estimator (MLE) and a memory non-linear estimator (MNLE): Starting with iteration count $t=1$, $\bm{x}_1 = {\rm E}\{\bm{x}\} = \bm{0}$,
\BS \label{Eqn:gen_MAMP}
\begin{alignat}{2}
    {\rm MLE:} && \quad \bm{r}_t &= \gamma_t(\bm{X}_t) = \bm{Q}_t\bm{y} + \textstyle\sum_{i=1}^t{\bm{P}}_{t,i} \bm{x}_i, \label{Eqn:gen_MLE} \\
    {\rm MNLE:} && \quad \bm{x}_{t+1} &= \phi_t(\bm{R}_t), \label{Eqn:gen_MNLE}
\end{alignat}
\ES
where $\bm{X}_t = [\bm{x}_1\, \cdots\, \bm{x}_t]$, $\bm{R}_t = [\bm{r}_1\, \cdots\, \bm{r}_t]$, and more details are as follows:
\begin{itemize}
    \item Let $\bm{f}_t = \bm{x}_t - \bm{x}$ and $\bm{g}_t = \bm{r}_t - \bm{x}$ indicate the errors. The following orthogonality constraints hold:
    for $k \leq t$,
    \BS\begin{align}
        &\lim_{N\to\infty} \tfrac{1}{N} \bm{g}_t^{\rm H}\bm{f}_{k} \overset{\rm a.s.}{=} 0, \label{Eqn:orth_1} \\
        &\lim_{N\to\infty} \tfrac{1}{N} \bm{g}_t^{\rm H}\bm{x} \overset{\rm a.s.}{=} 0, \label{Eqn:orth_2} \\
        &\lim_{N\to\infty} \tfrac{1}{N} \bm{f}_{t+1}^{\rm H}\bm{g}_{k} \overset{\rm a.s.}{=} 0. \label{Eqn:orth_3}
    \end{align}\ES
    An MLE is said to be orthogonal if (\ref{Eqn:orth_1}) and (\ref{Eqn:orth_2}) hold, and an MNLE is said to be orthogonal if (\ref{Eqn:orth_3}) holds. In other words, an MAMP consists of an orthogonal MLE and an orthogonal MNLE.
    \item $\bm{Q}_t\bm{A}$ and $\bm{P}_{t,i}$ are polynomials in $\bm{A}^{\rm H} \bm{A}$, i.e., $\bm{Q}_t\bm{A} = \textstyle\sum_{k\geq1}\alpha_{t,k}(\bm{A}^{\rm H}\bm{A})^k$, $\bm{P}_{t,i} = \textstyle\sum_{k\geq0}\beta_{t,k}(\bm{A}^{\rm H}\bm{A})^k$, where $\alpha_{t,k}$ and $\beta_{t,k}$ are real-valued. Without loss of generality, we assume that the norms of $\bm{Q}_t$ and $\bm{P}_{t,i}$ are finite.
\end{itemize}
\end{definition}
In addition, we define the covariance for $\bm{x}_i$ and $\bm{x}_j$ ($\bm{r}_i$ and $\bm{r}_j$) as
\BS\begin{align}
    v^{{\phi}}_{i,j} &\equiv \tfrac{1}{N}{\mr E} \{\bm{f}_{i}^{\mr H} \bm{f}_{j}\}, \\
    v^{\gamma}_{i,j} &\equiv \tfrac{1}{N}{\mr E} \{\bm{g}_{i}^{\mr H} \bm{g}_{j}\}.
\end{align}\ES

\subsection{Gradient Descent MAMP (GD-MAMP)}
\textbf{\emph{GD-MAMP Algorithm}}: Let ${\lambda}^\dag = ( \lambda_{\max} + \lambda_{\min}) / 2$ and $\bm{B} = \lambda^\dag\bm{I} - \bm{A}\bm{A}^{\mr H}$. GD-MAMP is an MAMP algorithm: Starting with $t=1$, $\bm{u}_{0} = \bm{0}_M$, $\bm{x}_1={\rm E}\{\bm{x}\}=\bm{0}_N$, 
\BS\label{Eqn:GD-MAMP}
\begin{alignat}{2}
    {\rm MLE:} && \bm{u}_{t} &= \theta_t \bm{B} \bm{u}_{t-1} + \xi_t(\bm{y} - \bm{A}\bm{x}_t), \label{Eqn:MLE1}\\
    && \bm{r}_t &=\gamma_t (\bm{X}_t) = \tfrac{1}{{\varepsilon}^\gamma_t}\big( \bm{A}^{\mr H}\bm{u}_{t} + \textstyle\sum_{i=1}^t  p_{t, i}\bm{x}_i \big), \label{Eqn:MLE2}\\
    {\rm NLE:} && \ \bm{x}_{t + 1} &= \bar\phi_t(\bm{r}_t) = \big[\bm{x}_1\,\cdots\,\bm{x}_t\; \phi_t ( \bm{r}_t)\big] \cdot \scaleto{\bm{\zeta}}{8pt}_{t+1}. \label{Eqn:NLE}
\end{alignat}\ES
The detailed parameters of the MLE are as follows:
\begin{enumerate}
    \item
    Let $\rho_t = \sigma^2 / v_{t,t}^{\bar{\phi}}$, where $v_{t,t}^{\bar{\phi}}$ for $t \geq 2$ is given by (\ref{Eqn:v_phi_bar}) and $v_{1,1}^{\bar{\phi}} = (\tfrac{1}{N} \|\bm{y}-\bm{A}\bm{x}_1\|^2 - \delta \sigma^2)/w_0$. The relaxation parameter $\theta_t$ is optimized by
    \begin{align}\label{Eqn:theta}
        \theta_t = (\lambda^\dag + \rho_t)^{-1}.
    \end{align}
    \item{
    For $i \geq 0$, $j \geq 0$, 
    \BS\begin{align}
        b_i &\equiv \tfrac{1}{N}{\rm tr}\{\bm{B}^i\} = \textstyle\sum_{i=0}^{t} \binom{t}{i} (-1)^i(\lambda^\dag)^{t-i}\lambda_i, \label{Eqn:b}\\
        w_i &\equiv  \tfrac{1}{N}{\rm tr}\{\bm{A}^{\rm H}\bm{B}^{i}\bm{A}\} = \lambda^\dag b_i - b_{i+1}, \label{Eqn:w}\\
        \bar{w}_{i,j} &\equiv \lambda^\dag  w_{i+j}-w_{i+j+1}-w_{i}w_{j}.
    \end{align}\ES
    Let
    \begin{align}\label{Eqn:vartheta}
        \vartheta_{t, i} = \xi_i \textstyle\prod_{\tau=i+1}^t\theta_\tau,
    \end{align}
    where $\xi_1 = 1$, and $\xi_t$ for $t > 1$ is optimized by  
    \begin{align}\label{Eqn:xi}
        \xi_t = \dfrac{c_{t,2}c_{t,0}+c_{t,3}}{c_{t,1}c_{t,0}+c_{t,2}},
    \end{align}
    with $c_{t, 0}, \cdots\!, c_{t, 3}$ given in \cite{liu2022memory}. The normalization parameter ${\varepsilon}^\gamma_t$ and orthogonalization parameters $\{p_{t,i}\}$ are given by
   \BS
   \begin{align}
        p_{t, i} &= \vartheta_{t, i} w_{t-i}, \label{Eqn:p_ti}\\ 
        {\varepsilon}^\gamma_t &= \textstyle\sum_{i=1}^t p_{t, i}.
    \end{align}
    \ES
    }
    \item
    For $1\leq t'\leq t$, $v_{t,t'}^{\gamma}$ is given by:
    \begin{align}\label{Eqn:v_gam}
        \begin{split}
            v^{\gamma}_{t,t'} = \tfrac{1}{\varepsilon^\gamma_{t}\varepsilon^\gamma_{t'}} \textstyle{\sum}_{i=1}^{t}\textstyle{\sum}_{j=1}^{t'}  \vartheta_{t, i}\vartheta_{t'\!,j}\big[\sigma^2  w_{t+t'-i-j} \\
            + v^{\bar{\phi}}_{i,j}\bar{ w}_{t-i,t'\!-j}\big].
        \end{split}
    \end{align}
\end{enumerate}
The detailed parameters of the NLE are as follows:
\begin{enumerate}
    \item{
    $\phi_t(\cdot)$ is a separable and Lipschitz-continuous function, which is the same as that in OAMP/VAMP \cite{ma2017orthogonal}, \cite{rangan2019vector}. 
    }
    \item{
    For $t \geq 2$,
    \begin{align}\label{Eqn:v_phi}
    v^{\phi}_{t,k} \overset{\rm a.s.}{=}
    \begin{cases}
        \lim\limits_{N\to\infty}(\tfrac{1}{N} \|\bar{\bm{z}}_t\|^2 - \delta \sigma^2)/w_0, & k = t \ \\[1mm]
        \lim\limits_{N\to\infty}(\tfrac{1}{N} \bar{\bm{z}}_t^{\mr H} \bm{z}_{k} - \delta \sigma^2)/w_0 , & k < t 
    \end{cases}.
    \end{align}
    where $\bar{\bm{z}}_t = \bm{y}-\bm{A}\,\phi_{t-1}(\bm{r}_{t-1})$ and $\bm{z}_{k} = \bm{y}-\bm{A}\bm{x}_{k}$. As a algorithmic step, we just replace $\overset{\rm a.s.}{=}$ with $=$, and remove the symbol ``lim''.
    }
    \item{
    Let $\bm{V}_{t+1}^{\phi}$ be the covariance matrix for $\bm{x}_1, \cdots\!, \bm{x}_{t}, \phi_t(\bm{r}_t)$, which is assumed to be invertible. The damping vector $\scaleto{\bm{\zeta}}{8pt}_{t+1}$ is given by 
    \begin{align}\label{Eqn:xi_2}
        \scaleto{\bm{\zeta}}{8pt}_{t+1} = 
        \frac{(\bm{V}_{t+1}^{\phi})^{-1} \bm{1}}{\bm{1}^{\rm T} (\bm{V}_{t+1}^{\phi})^{-1}\bm{1}}.
    \end{align}
    Let $\bm{v}^{\bar{\phi}}_{t+1} = [v^{\bar{\phi}}_{t+1,1} \,\cdots\, v^{\bar{\phi}}_{t+1,t+1}]^{\rm T}$, we have
    \begin{align}\label{Eqn:v_phi_bar}
        \bm{v}^{\bar{\phi}}_{t+1} = \dfrac{\bm{1}}{\bm{1}^{\rm T} (\bm{V}_{t+1}^{\phi})^{-1}\bm{1}}.
    \end{align}
    Hence, $v^{\bar{\phi}}_{t+1,k} = v^{\bar{\phi}}_{k,t+1} = v^{\bar{\phi}}_{t+1,t+1}$ holds for $1 \leq k \leq t$. In practice, the maximum damping length is generally set as $L$ to improve robustness and reduce complexity, where $L\ll T$. If $\bm{V}_{t+1}^{\phi}$ is singular, we employ a back-off damping strategy. Further details can be found in \cite{liu2022memory}.
    }
\end{enumerate}

\subsection{Simplification of GD-MAMP}\label{SSec:SGD}
The main complexity of GD-MAMP is $\mathcal{O}(MNT)$, which is dominated by the number of matrix-vector products. For $t > 1$, GD-MAMP requires four matrix-vector products per iteration, outlined as follows\footnote{Note that $\bm{z}_t = \bm{y} - \bm{A}\bm{x}_t = [\bm{z}_1\,\cdots\,\bm{z}_{t-1}\; \bar{\bm{z}}_t] \cdot \scaleto{\bm{\zeta}}{6pt}_{t+1}$. Hence, computing $\bm{z}_t$ does not a require matrix-vector product.}: 
\begin{itemize}
    \item Computing $\bar{\bm{z}}_t = \bm{y}-\bm{A}\phi_{t-1}(\bm{r}_{t-1})$ requires one.
    \item Computing $\bm{B}\bm{u}_{t-1} = (\lambda^\dag\bm{I} - \bm{A}\bm{A}^{\mr H})\bm{u}_{t-1}$ in (\ref{Eqn:MLE1}) requires two.
    \item Computing $\bm{A}^{\mr H}\bm{u}_{t}$ in (\ref{Eqn:MLE2}) requires one.
\end{itemize}
The computation of $\bm{A}^{\rm H}\bm{u}_{t-1}$ is redundant, as it was calculated in the $(t\!-\!1)$-th iteration. Thus, we rewrite (\ref{Eqn:GD-MAMP}) as: Starting with $t = 1$, $\bm{x}_1 = {\rm E}\{\bm{x}\} = \bm{0}_{N}$, $\bm{u}_{0} = \bm{0}_{M}$ and $\hat{\bm{r}}_{0} = \bm{0}_{N}$,
\BS
\label{Eqn:SGD_MAMP}
\begin{alignat}{2}
    {\rm MLE:} && \bm{u}_{t} &= \theta_t\lambda^\dag\bm{u}_{t-1} + \xi_t\bm{y} - \bm{A}(\theta_t\hat{\bm{r}}_{t-1} + \xi_t\bm{x}_t), \label{Eqn:SGD_MAMP_a}\\
    && \hat{\bm{r}}_{t} &= \bm{A}^{\mr H}\bm{u}_{t}, \label{Eqn:SGD_MAMP_b}\\
    && \bm{r}_t &= \gamma_t(\bm{X}_t) = \tfrac{1}{{\varepsilon}^\gamma_t}\big(\hat{\bm{r}}_{t} + \textstyle\sum_{i=1}^t  p_{t, i}\bm{x}_i \big), \label{Eqn:SGD_MAMP_c}\\
    {\rm NLE:} &&\ \bm{x}_{t+1} &= \bar\phi_t(\bm{r}_t) = \big[\bm{x}_1\,\cdots\,\bm{x}_t\; \phi_t ( \bm{r}_t)\big] \cdot \scaleto{\bm{\zeta}}{8pt}_{t+1}. 
\end{alignat}
\ES
As a result, the number of matrix-vector products required in each iteration of GD-MAMP can be decreased from four to three, leading to an almost $25\%$ reduction of the time cost. 

\section{Overflow-Avoiding GD-MAMP}\label{Sec:OA}
In this section, we conduct an analysis of the potential overflow problem in GD-MAMP. Then, we propose an overflow-avoiding GD-MAMP (OA-GD-MAMP) algorithm, which has two different forms depending on whether the eigenvalues of $\bm{A}\bm{A}^{\rm H}$ are known.

\subsection{Overflow Problem}
In GD-MAMP, it is necessary to compute $\{w_t\}$ for $t < 2T$, where $w_t = \lambda^\dag b_i - b_{i+1}$ and $b_i = \tfrac{1}{N}{\rm tr}\{\bm{B}^i\}$. Thus, we need to compute $\{b_t\}$ for $t \leq 2T$. However, for $k \leq t$, when the spectral radius $\rho(\bm{B}) > 1$, $b_{2k}$ increases exponentially as $k$ increases. Note that when matrix $\bm{A}$ is highly ill-conditioned, the convergence of GD-MAMP tends to be slow, thereby requiring a larger $T$. In such cases, GD-MAMP fails to operate due to the inability to store the extremely large value of $b_{2k}$. We refer to this problem as overflow problem of GD-MAMP.

\subsection{OA-GD-MAMP with Eigenvalues of $\bm{A}\bm{A}^{\rm H}$}
First, we consider that the eigenvalues of $\bm{A}\bm{A}^{\rm H}$ are known. Let $\bm{\lambda}_{\bm{B}}$ denote the vector of the eigenvalues of $\bm{B} = \lambda^\dag\bm{I} - \bm{A}\bm{A}^{\mr H}$. For $\bm{a}=[a_1, \cdots\!, a_k]^{\rm T}$, $\bm{b}=[b_1, \cdots\!, b_k]^{\rm T}$, we define $|\bm{a}| \equiv \big[|a_1|, \cdots\!, |a_k| \big]^{\rm T}$, ${\rm sgn}(\bm{a}) \equiv \big[{\rm sgn}(a_1),\cdots\!, {\rm sgn}(a_k)\big]^{\rm T}$, $\log^{\circ}(\bm{a}) \equiv \big[\log(a_1), \cdots\!, \log(a_k)\big]^{\rm T}$, $\mathrm{e}^{\circ \bm{a}} \equiv [\mathrm{e}^{a_1}, \cdots\!, \mathrm{e}^{a_k}]^{\rm T}$, $\bm{a} \circ \bm{b} \equiv [a_1 b_1, \cdots\!, a_k b_k]^{\rm T}$, and $\bm{a}^{\circ c} \equiv [a_1^c, \cdots\!, a_k^c], c \in \mathbb{C}$.

In the $t$-th iteration of GD-MAMP, the terms like $\vartheta_{t,i}w_{t-i}$ and $\vartheta_{t, i}\vartheta_{t,j}w_{2t-i-j}$ for $i,j = 1, \cdots\!, t$ need to be calculated. When $w_{t-i}$ or $w_{2t-i-j}$ is very large, $\vartheta_{t,i}$ or $\vartheta_{t, i}\vartheta_{t,j}$ is very close to $0$. Due to the following Lemma \ref{Lemma:oa_1}, we can compute these terms without explicitly calculating $w_{t-i}$ or $w_{2t-i-j}$ to avoid the overflow problem, by utilizing $\bm{\lambda}_{\bm{B}}$.
\begin{lemma}\label{Lemma:oa_1}
    For any $\alpha \in \mathbb{R}\backslash\{0\}$, $k \geq 0$,
    \BS\begin{align}
        \alpha w_k &= \frac{{\rm sgn}(\alpha)}{N} \bm{1}^{\rm T} \big[(\lambda^\dag\bm{1}-\bm{\lambda_B}) \circ \bm{s}_\lambda^{\circ k} \circ \mathrm{e}^{\circ \log|\alpha|\bm{1}+k\bm{\lambda}_{B}^{\log}} \big], 
    \end{align}
    where
    \begin{align}
        \bm{s}_{\lambda} &\equiv {\rm sgn}(\bm{\lambda}_{B}), \\
        \bm{\lambda}_{B}^{\log} &\equiv \log^{\circ}|\bm{\lambda}_{B}|.
    \end{align}\ES
\end{lemma}
\begin{IEEEproof}
    Recall $w_k = \lambda^\dag b_k - b_{k+1}$, where $b_k = \tfrac{1}{N}{\rm tr}\{\bm{B}^k\}$. 
    \begin{align}
    \alpha w_k &= \tfrac{1}{N} \alpha \big(\lambda^\dag{\rm tr}\{\bm{B}^k\} - {\rm tr}\{\bm{B}^{k+1}\}\big) \nonumber \\
    &= \tfrac{1}{N} \alpha \bm{1}^{\rm T} \big[(\lambda^\dag\bm{1}-\bm{\lambda}_{B})   \circ \bm{\lambda_B}^{\circ k} \big]. \label{Eqn:L3_p1}
    \end{align}
    Notice that 
    \BS\label{Eqn:L3_p2}\begin{align}
        \alpha &= {\rm sgn}(\alpha) \mathrm{e}^{\log|\alpha|}, \\
        \bm{\lambda_B}^{\circ k} &= \bm{s}_{\lambda}^{\circ k}  \circ \mathrm{e}^{\circ k\bm{\lambda}_{B}^{\log}}.
    \end{align}\ES
    By substituting (\ref{Eqn:L3_p2}) into (\ref{Eqn:L3_p1}), we completed the proof.
\end{IEEEproof}
However, applying Lemma \ref{Lemma:oa_1} requires $\mathcal{O}(M)$, and GD-MAMP requires to compute $\mathcal{O}(T^3)$ terms involving $w_k$. If using Lemma \ref{Lemma:oa_1} to compute these terms each time, the overall complexity is excessively high as $\mathcal{O}(MT^3)$. As an alternative, we construct 
\begin{align}\label{Eqn:rchi}
    \rchi_{k} \equiv \theta_0^k w_k,
\end{align}
where $\theta_0 = (\lambda^\dag + \sigma^2)^{-1}$. Thus, we can pre-calculate $\{\rchi_{k}\}$ for $k < 2T$ by Lemma \ref{Lemma:oa_1} before the iterations begin, and obtain each term involving $w_k$ by
\begin{align}\label{Eqn:bw}
    \alpha w_k = {\rm sgn}(\alpha)\mathrm{e}^{\log|\alpha|-k\log\theta_0}\rchi_k.
\end{align}
As a result, the complexity becomes $\mathcal{O}(MT+T^3)$. The following Lemma \ref{Lemma:oa_2} shows a bound of $|\rchi_{k}|$, ensuring that computing $\{\rchi_{k}\}$ has no risk of overflow.
\begin{lemma}\label{Lemma:oa_2}
    For any $k\geq 0$, 
    \begin{align}
        |\rchi_{k}| \leq \delta (\lambda^\dag + \theta_0^{-1}),
    \end{align}
    where $\delta = M/N$.
\end{lemma}
\begin{IEEEproof}
    Recall $w_k = \lambda^\dag b_k - b_{k+1}$, where $b_k = \tfrac{1}{N}{\rm tr}\{\bm{B}^k\}$. We have
    \begin{align}\label{Eqn:Xi}
        |\rchi_{k}| = \big| \lambda^\dag \theta_0^k b_k - \theta_0^{-1} \theta_0^{k+1} b_{k+1} \big|.
    \end{align}
    Since the spectral radius of $\bm{B}$ is $\lambda^{\dag}-\lambda_{\min}$, the spectral radius of $\theta_0\bm{B}$ is less than $1$. For $k \geq 0$,
    \begin{align}\label{Eqn:theta_b}
        -\delta \leq \theta_0^k b_k = \tfrac{1}{N}{\rm tr}\{(\theta_0\bm{B})^k\} \leq \delta.
    \end{align} 
   Both $\lambda^\dag$ and $\theta_0$ are positive. Thus, following \eqref{Eqn:Xi} and \eqref{Eqn:theta_b}, we have $|\rchi_{k}| \leq \delta (\lambda^\dag + \theta_0^{-1})$.
\end{IEEEproof}

We refer to this equivalently reconstructed algorithm as overflow-avoiding GD-MAMP (OA-GD-MAMP).

\subsection{OA-GD-MAMP without Eigenvalues of $\bm{A}\bm{A}^{\rm H}$}
Now consider the scenario where the eigenvalues of $\bm{A}\bm{A}^{\rm H}$ are unknown. The first step is to approximate $\lambda_{\max}$ and $\lambda_{\min}$ by the method in \cite{liu2022memory}. Second, instead of calculating the sequence $\{\rchi_k\}$ by Lemma \ref{Lemma:oa_1}, we approximate $\{\rchi_k\}$ by the following lemma \ref{Lemma:oa_h}.
\begin{lemma}\label{Lemma:oa_h}
    For $k \geq 0$, 
    \begin{align}
        \rchi_k \overset{\rm a.s.}{=} \lim_{N\to\infty} \bar{\bm{h}}_i^{\rm H} \bar{\bm{h}}_{k-i} \label{Eqn:oa_rchi}
    \end{align}
    where $i = \lceil k/2 \rceil$ and $\bar{\bm{h}}_i$ is given by a recursion
    \begin{align}
        \bar{\bm{h}}_i = \theta_0\bm{B}\bar{\bm{h}}_{i-1},
    \end{align}
    with $\bar{\bm{h}}_0 = \bm{A}\bm{h}_0$, $\bm{h}_0 \sim \mathcal{N}(\bm{0},\tfrac{1}{N}\bm{I}_{N})$.
\end{lemma}
\begin{IEEEproof}
    For $0 \leq i \leq k$, 
    \BS
    \begin{align}
        \lim_{N\to\infty} \bar{\bm{h}}_i^{\rm H} \bar{\bm{h}}_{k-i} &= \lim_{N\to\infty} \bm{h}_0^{\rm H} \bm{A}^{\rm H}(\theta_0\bm{B})^k\bm{A}\bm{h}_0 \\ 
        &\overset{\rm a.s.}{=} \lim_{N\to\infty}\tfrac{\theta_0^k}{N}{\rm tr}\{\bm{A}^{\rm H}\bm{B}^k\bm{A}\} \bm{h}_0^{\rm H} \bm{h}_0 \\
        &= \theta_0^k w_k = \rchi_k.
    \end{align}
    Hence, we completed the proof.
    \ES
\end{IEEEproof}
Approximating $\lambda_{\max}$ and $\lambda_{\min}$ introduces $\tau$ matrix-vector products, where $\tau$ is usually set from $20$ to $30$. Approximating $\{\rchi_k\}$ for $k < 2T$ by Lemma \ref{Lemma:oa_h} introduces $T$ matrix-vector products. In other words, OA-GD-MAMP without eigenvalues of $\bm{A}\bm{A}^{\rm H}$ requires a total of $4T + \tau$ matrix-vector products.

\section{Complexity-Reduced GD-MAMP}
As shown in Subsection \ref{SSec:SGD}, GD-MAMP requires three matrix-vector products per iteration. In this section, we propose a complexity-reduced GD-MAMP (CR-GD-MAMP), reducing the number of matrix-vector products to two while maintaining comparable convergence speed.

\textbf{\emph{CR-GD-MAMP Algorithm}}: Starting with $t=1$, $\bm{x}_1 = {\rm E}\{\bm{x}\} = \bm{0}_{N}$ and $\bm{u}_{0} = \bm{0}_{M}$ and $\hat{\bm{r}}_{0} = \bm{0}_{N}$,
\BS\label{Eqn:MLE_CR}
\begin{alignat}{2}
    {\rm MLE:} && \bm{u}_{t} &= \theta_t\lambda^\dag\bm{u}_{t-1} + \tilde{\xi}_t\bm{y} - \bm{A}(\theta_t\hat{\bm{r}}_{t-1} + \tilde{\xi}_t\bm{x}_t), \\
    && \hat{\bm{r}}_{t} &= \bm{A}^{\mr H}\bm{u}_{t}, \\
    && \bm{h}_t &= \gamma_t(\bm{X}_t) = \tfrac{1} {{\varepsilon}^\gamma_t}\big(\hat{\bm{r}}_{t} + \textstyle\sum_{i=1}^t  p_{t, i}\bm{x}_i \big), \\
    && \bm{r}_t &= \bar{\gamma}_t(\bm{X}_t) = \big[\bm{h}_1 \cdots \bm{h}_t\big] \cdot \tilde{\scaleto{\bm{\zeta}}{8pt}}_t,
    \label{Eqn:MLE_CR_d}\\
    {\rm NLE:} &&\ \bm{x}_{t + 1} &= \phi_t(\bm{r}_t).
\end{alignat}\ES
where most parameters remain consistent with those of GD-MAMP. The differences are as follows: 
\begin{enumerate}
    \item The covariance matrix $\bm{V}_t^{\phi} \equiv [v_{i,j}^{\phi}]_{t \times t}$ are unknown now. We obtain $v_{t,t}^{\phi}$ as that in OAMP/VAMP:
    \begin{align}
        v_{t,t}^{\phi} = (1 / v_{t-1}^{\rm post} - 1 / v_{t-1,t-1}^{\bar{\gamma}})^{-1},
    \end{align}
    where $v_{t-1}^{\rm post}$ is the posterior variance given by the local MMSE demodulator in the NLE, and $v_{t-1, t-1}^{\bar{\gamma}}$ is the variance of $\bm{r}_{t-1}$ determined by the MLE. 
    \item Since $\bm{V}_t^{\phi}$ is unknown, we replace $\xi_t$ in (\ref{Eqn:xi}) with 
    \begin{align}
        \tilde{\xi}_t = 1 / (v_{t,t}^{\phi} + \sigma^2).
    \end{align}
    Though $\tilde{\xi}_t$ is sub-optimal, our simulation results demonstrate that it has little to no impact on the convergence speed. For some cases, even in GD-MAMP, $\tilde{\xi}_t$ is more robust than $\xi_t$. 
    \item We can no longer obtain the variance $v_{t,t}^{\gamma}$ of $\bm{h}_t$ by (\ref{Eqn:v_gam}) as $\bm{V}_t^{\phi}$ is unknown. Instead, we have
    \begin{align}
        v_{t, t}^{\gamma} \overset{\rm a.s.}{=} \lim_{N\to\infty}\tfrac{1}{N}\| \bm{h}_t - \bm{x}_t \|^2 - v_{t,t}^{\phi}.
    \end{align}
    This method was first utilized in \cite{skuratovs2022compressed}. In addition, 
    the following Proposition \ref{Prop:CR_cov} indicate how to obtain $\Re(v_{t, i}^{\gamma})$.
    \begin{proposition}\label{Prop:CR_cov}
        For $i \leq t$, we have 
        \begin{align}
            \Re(v_{t, i}^{\gamma}) \overset{\rm a.s.}{=} \lim_{N\to\infty} \big(v_{t, t}^{\gamma} + v_{i, i}^{\gamma} - \tfrac{1}{N}\|\bm{h}_{t} - \bm{h}_i\|^2 \big) / 2. 
        \end{align}
    \end{proposition}
    \begin{IEEEproof}
        Let $\bm{h}_t = \bm{x} + \hat{\bm{g}}_t$. We have
        \begin{align}
            &\lim_{N\to\infty} \tfrac{1}{N}\|\bm{h}_{t} - \bm{h}_i\|^2 \nonumber \\
            &= \lim_{N\to\infty} \tfrac{1}{N}(\hat{\bm{g}}_t^{\rm H}\hat{\bm{g}}_t + \hat{\bm{g}}_i^{\rm H}\hat{\bm{g}}_i - \hat{\bm{g}}_i^{\rm H}\hat{\bm{g}}_t - \hat{\bm{g}}_t^{\rm H}\hat{\bm{g}}_i) \nonumber \\
            &\overset{\rm a.s.}{=} v_{t, t}^{\gamma} + v_{i, i}^{\gamma} - 2\Re(v_{t, i}^{\gamma}).
        \end{align}
        Thus, we finished the proof.
    \end{IEEEproof}
    \begin{figure}[t] 
    \centering
    \includegraphics[width=63mm]{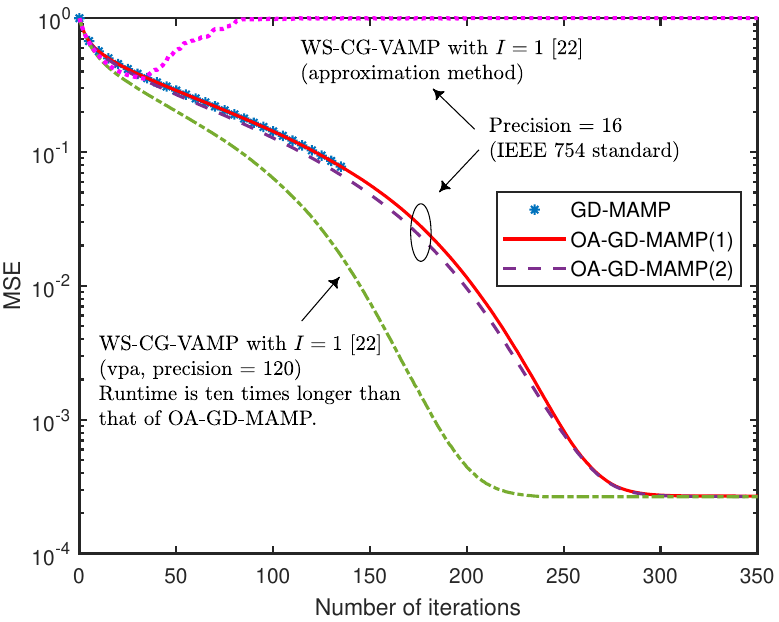} \\
    \caption{MSE versus the number of iterations. A real-valued system with $M=8192, N=16384$, $\mu=0.1$, ${\rm SNR}=35$dB, $\kappa = 1000$. OA-GD-MAMP(1)/(2) denotes OA-GD-MAMP with/without eigenvalues of $\bm{A}\bm{A}^{\rm H}$.}
    \label{Fig:OA}
    \end{figure}

    Suppose that $\bm{V}_t^{\gamma} \equiv [v_{i,j}^{\gamma}]_{t \times t}$ is invertible\footnote{If $V_t^{\gamma}$ is singular, we employ the same back-off damping strategy as in GD-MAMP. }. Then, the damping vector $\tilde{\scaleto{\bm{\zeta}}{8pt}}_t$ is given by
    \begin{align}\label{Eqn:CR_da}
        \tilde{\scaleto{\bm{\zeta}}{8pt}}_t = 
        \frac{[\Re(\bm{V}_{t}^{\gamma})]^{-1} \bm{1}}{\bm{1}^{\rm T} [\Re(\bm{V}_{t}^{\gamma})]^{-1}\bm{1}}.
    \end{align}
    For real-valued systems, i.e, $\bm{A}\in \mathbb{R}^{M \times N}$, $\bm{x}\in \mathbb{R}^N$, $\tilde{\scaleto{\bm{\zeta}}{8pt}}_t$ in (\ref{Eqn:CR_da}) is the optimal damping as proven in \cite{liu2022memory}, while for complex-valued systems, it remains optimal if we constrain the damping vector to be real-valued.
    \begin{proposition}\label{Prop:CR2}
        Suppose that $\bm{V}_t^{\gamma}$ is invertible and the damping vector is real-valued. Then, $\tilde{\scaleto{\bm{\zeta}}{8pt}}_t$ in (\ref{Eqn:CR_da}) is the optimal damping vector that minimizes the variance $v_{t, t}^{\bar{\gamma}}$.
    \end{proposition}
    \begin{IEEEproof}
        $\tilde{\scaleto{\bm{\zeta}}{8pt}}_t$ proves to be the solution of 
        \begin{align}
            & \min_{\scaleto{\bm{\zeta}}{8pt}_t}\ \; \scaleto{\bm{\zeta}}{8pt}_t^{\rm H}\,\Re(\bm{V}_t^{\gamma})\, \scaleto{\bm{\zeta}}{8pt}_t, \quad {\rm s.t.}\ \bm{1}^{\rm T}\scaleto{\bm{\zeta}}{8pt}_t = 1.
        \end{align}
        Notice that $\scaleto{\bm{\zeta}}{8pt}_t^{\rm H}\, \bm{V}_t^{\gamma}\, \scaleto{\bm{\zeta}}{8pt}_t = \scaleto{\bm{\zeta}}{8pt}_t^{\rm H}\, \Re(\bm{V}_t^{\gamma}) \,\scaleto{\bm{\zeta}}{8pt}_t$ if $\scaleto{\bm{\zeta}}{8pt}_t \in \mathbb{R}^t$. Since $\tilde{\scaleto{\bm{\zeta}}{8pt}}_t \in \mathbb{R}^t$, it is the solution of
        \begin{align}
            & \min_{\scaleto{\bm{\zeta}}{8pt}_t}\ \;\scaleto{\bm{\zeta}}{8pt}_t^{\rm H}\,\bm{V}_t^{\gamma}\,\scaleto{\bm{\zeta}}{8pt}_t,\quad {\rm s.t.}\ \bm{1}^{\rm T}\scaleto{\bm{\zeta}}{8pt}_t = 1,\ \scaleto{\bm{\zeta}}{8pt}_t \in \mathbb{R}^t.
        \end{align}
        Thus, we have finished the proof.
    \end{IEEEproof}
    Finally, $v^{\bar{\gamma}}_{t, t}$ is given by
    \begin{align}
        v^{\bar{\gamma}}_{t, t} = \dfrac{1}{\bm{1}^{\rm T} [\Re(\bm{V}_t^{\gamma})]^{-1}\bm{1}}.
    \end{align}
    In practice, similar to GD-MAMP, we set the maximum damping length as $L$, where $L \ll T$. 
\end{enumerate}

\section{Simulation Results}
\subsection{Experiment Settings}
The signal $\bm{x} = [x_1, \cdots\!, x_N]^{\rm T}$ is IID. For real-valued systems, each element $x_k$ follows a Bernoulli-Gaussian distribution, i.e., for $1 \leq k \leq N$,
\begin{align}\label{Eqn:BG}
    x_k = 
    \begin{cases}
    0, & \text{probability} = 1 - \mu \\[1mm]
    {\cal N}(0, \tfrac{1}{\mu}), & \text{probability} = \mu
    \end{cases}.
\end{align}
For complex-valued systems, we have $x_k = (x_k^{Re} + x_k^{Im}) / \sqrt{2}$, where $x_k^{Re}$ and $x_k^{Im}$ are independently given by (\ref{Eqn:BG}).

The matrix $\bm{A}$ is constructed as 
\begin{align}
    \bm{A} = \bm{\Sigma}\bm{\Pi}\bm{F},
\end{align}
where $\bm{\Pi}$ is an $N \times N$ random permutation matrix, $\bm{F}$ is the normalized discrete cosine transform (DCT) or discrete Fourier transform (DFT) matrix. Let $J = \min(M, N)$. $\bm{\Sigma} \in \mathbb{R}^{M \times N}$ is a diagonal matrix with diagonal entries $\{\sigma_i\}$ for $1 \leq i < J$, such that $\sigma_i/\sigma_{i+1} = \kappa^{1/J}$ and $\tfrac{1}{N}\textstyle\sum_{i=1}^{J}\sigma_i^2 = 1$. Under this construction, the condition number of $\bm{A}$ is $\kappa^{1-1/J}$, which tends to $\kappa$ when $J$ is large.

\subsection{Numerical Simulations}
Fig.~\ref{Fig:OA} shows that GD-MAMP fails to operate when the number of iterations $t > 136$ due to the overflow problem. In contrast, OA-GD-MAMP solves the overflow problem and works well. In addition, OA-GD-MAMP without eigenvalues of $\bm{A}\bm{A}^{\rm H}$ achieves comparable performance. On the other hand, in support of our comments about WS-CG-VAMP in Section \ref{Sec:background}, we simulate WS-CG-VAMP with number of inner iterations $I=1$. The result shows that WS-CG-VAMP works properly with MATLAB variable-precision arithmetic (vpa) of 120. However, employing vpa makes the runtime of WS-CG-VAMP several tens of times greater than that of GD-MAMP. Importantly, the approximation method in \cite{skuratovs2022compressed} is not effective in this scenario, causing WS-CG-VAMP to diverge. In other words, the precision problem of WS-CG-VAMP is still unsolved.
\begin{figure}[t] 
    \centering
    \includegraphics[width=63mm]{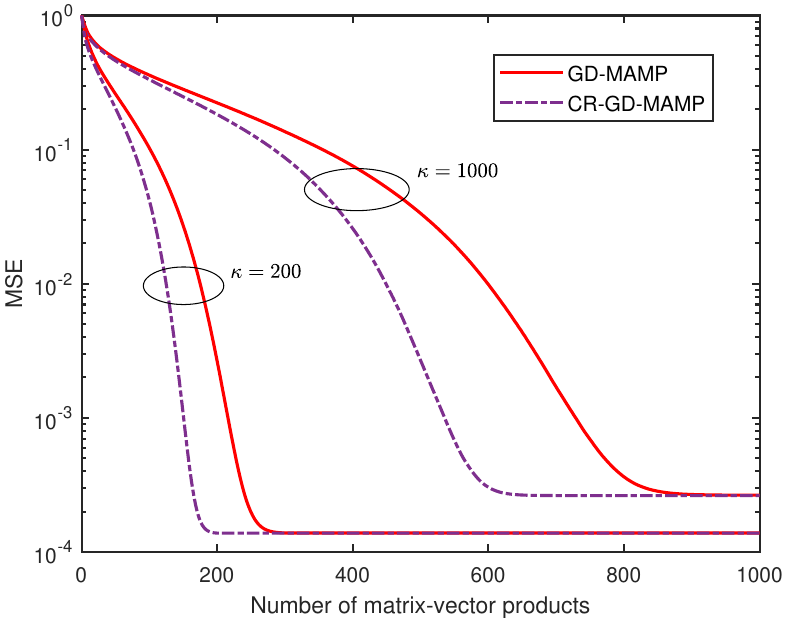} \\
    \caption{MSE versus the number of matrix-vector products. A complex-valued system with $M=8192, N=16384$, $\mu=0.1$, ${\rm SNR}=35$dB, $\kappa = 1000$.}
    \label{Fig:CR}
\end{figure}

In Fig.~\ref{Fig:CR}, to compare CR-GD-MAMP and GD-MAMP fairly, we focus on MSE versus the number of matrix-vector products instead of the number of iterations. As can be seen, CR-GD-MAMP converges to the same point as GD-MAMP while using about $2/3$ the number of matrix-vector products.

\section{Conclusion}
In this paper, we solved the overflow problem of GD-MAMP. Additionally, we developed a variant of GD-MAMP to further reduce the complexity. 





\bibliographystyle{IEEEtran}
\bibliography{reference}

\begin{thebibliography}{10}
\providecommand{\url}[1]{#1}
\csname url@samestyle\endcsname
\providecommand{\newblock}{\relax}
\providecommand{\bibinfo}[2]{#2}
\providecommand{\BIBentrySTDinterwordspacing}{\spaceskip=0pt\relax}
\providecommand{\BIBentryALTinterwordstretchfactor}{4}
\providecommand{\BIBentryALTinterwordspacing}{\spaceskip=\fontdimen2\font plus
\BIBentryALTinterwordstretchfactor\fontdimen3\font minus \fontdimen4\font\relax}
\providecommand{\BIBforeignlanguage}[2]{{%
\expandafter\ifx\csname l@#1\endcsname\relax
\typeout{** WARNING: IEEEtran.bst: No hyphenation pattern has been}%
\typeout{** loaded for the language `#1'. Using the pattern for}%
\typeout{** the default language instead.}%
\else
\language=\csname l@#1\endcsname
\fi
#2}}
\providecommand{\BIBdecl}{\relax}
\BIBdecl

\bibitem{verdu1985optimum}
S.~Verd{\'u}, ``Optimum multiuser signal detection ({P}h.{D}. {A}bstr.),'' \emph{{IEEE} Trans. Inf. Theory}, vol.~31, no.~4, pp. 557--557, Jul. 1985.

\bibitem{micciancio2001hardness}
D.~Micciancio, ``The hardness of the closest vector problem with preprocessing,'' \emph{{IEEE} Trans. Inf. Theory}, vol.~47, no.~3, pp. 1212--1215, Mar. 2001.

\bibitem{donoho2009message}
D.~L. Donoho, A.~Maleki, and A.~Montanari, ``Message-passing algorithms for compressed sensing,'' \emph{Proc. Natl. Acad. Sci. U.S.A.}, vol. 106, no.~45, pp. 18\,914--18\,919, Nov. 2009.

\bibitem{bayati2011dynamics}
M.~Bayati and A.~Montanari, ``The dynamics of message passing on dense graphs, with applications to compressed sensing,'' \emph{{IEEE} Trans. Inf. Theory}, vol.~57, no.~2, pp. 764--785, Feb. 2011.

\bibitem{reeves2019replica}
G.~Reeves and H.~D. Pfister, ``The replica-symmetric prediction for random linear estimation with {G}aussian matrices is exact,'' \emph{{IEEE} Trans. Inf. Theory}, vol.~65, no.~4, pp. 2252--2283, Apr. 2019.

\bibitem{barbier2020mutual}
J.~Barbier, N.~Macris, M.~Dia, and F.~Krzakala, ``Mutual information and optimality of approximate message-passing in random linear estimation,'' \emph{{IEEE} Trans. Inf. Theory}, vol.~66, no.~7, pp. 4270--4303, Jul. 2020.

\bibitem{liu2021capacity}
L.~Liu, C.~Liang, J.~Ma, and L.~Ping, ``Capacity optimality of {AMP} in coded systems,'' \emph{{IEEE} Trans. Inf. Theory}, vol.~67, no.~7, pp. 4429--4445, Jul. 2021.

\bibitem{manoel2014sparse}
A.~Manoel, F.~Krzakala, E.~W. Tramel, and L.~Zdeborov{\'a}, ``Sparse estimation with the swept approximated message-passing algorithm,'' \emph{arXiv preprint arXiv:1406.4311}, 2014.

\bibitem{vila2015adaptive}
J.~Vila, P.~Schniter, S.~Rangan, F.~Krzakala, and L.~Zdeborov{\'a}, ``Adaptive damping and mean removal for the generalized approximate message passing algorithm,'' in \emph{{IEEE} International Conference on Acoustics, Speech, and Signal Processing (ICASSP)}, 2015, pp. 2021--2025.

\bibitem{rangan2017inference}
S.~Rangan, A.~K. Fletcher, P.~Schniter, and U.~S. Kamilov, ``Inference for generalized linear models via alternating directions and bethe free energy minimization,'' \emph{{IEEE} Trans. Inf. Theory}, vol.~63, no.~1, pp. 676--697, Jan. 2017.

\bibitem{guo2015approximate}
Q.~Guo and J.~Xi, ``Approximate message passing with unitary transformation,'' \emph{arXiv preprint arXiv:1504.04799}, 2015.

\bibitem{ma2017orthogonal}
J.~Ma and L.~Ping, ``Orthogonal {AMP},'' \emph{{IEEE} Access}, vol.~5, pp. 2020--2033, Jan. 2017.

\bibitem{rangan2019vector}
S.~Rangan, P.~Schniter, and A.~K. Fletcher, ``Vector approximate message passing,'' \emph{{IEEE} Trans. Inf. Theory}, vol.~65, no.~10, pp. 6664--6684, May 2019.

\bibitem{takeuchi2020rigorous}
K.~Takeuchi, ``Rigorous dynamics of expectation-propagation-based signal recovery from unitarily invariant measurements,'' \emph{{IEEE} Trans. Inf. Theory}, vol.~66, no.~1, pp. 368--386, Jan. 2020.

\bibitem{liu2021capacity_oamp}
L.~Liu, S.~Liang, and L.~Ping, ``Capacity optimality of {OAMP}: Beyond iid sensing matrices and gaussian signaling,'' \emph{arXiv preprint arXiv:2108.08503}, 2021.

\bibitem{liu2023oamp}
L.~Liu, Y.~Cheng, S.~Liang, J.~H. Manton, and L.~Ping, ``On {OAMP}: Impact of the orthogonal principle,'' \emph{{IEEE} Trans. Commun.}, vol.~71, no.~5, pp. 2992--3007, May 2023.

\bibitem{takeuchi2019unified}
K.~Takeuchi, ``A unified framework of state evolution for message-passing algorithms,'' in \emph{{IEEE} International Symposium on Information Theory (ISIT)}, 2019, pp. 151--155.

\bibitem{fan2022approximate}
Z.~Fan, ``Approximate message passing algorithms for rotationally invariant matrices,'' \emph{Ann. Stat.}, vol.~50, no.~1, pp. 197--224, 2022.

\bibitem{dudeja2022spectral}
R.~Dudeja, S.~Sen, and Y.~M. Lu, ``Spectral universality of regularized linear regression with nearly deterministic sensing matrices,'' \emph{arXiv preprint arXiv:2208.02753}, 2022.

\bibitem{takeuchi2021bayes}
K.~Takeuchi, ``Bayes-optimal convolutional {AMP},'' \emph{{IEEE} Trans. Inf. Theory}, vol.~67, no.~7, pp. 4405--4428, May 2021.

\bibitem{liu2022memory}
L.~Liu, S.~Huang, and B.~M. Kurkoski, ``Memory {AMP},'' \emph{{IEEE} Trans. Inf. Theory}, vol.~68, no.~12, pp. 8015--8039, Jun. 2022.

\bibitem{skuratovs2022compressed}
N.~Skuratovs and M.~E. Davies, ``Compressed sensing with upscaled vector approximate message passing,'' \emph{{IEEE} Trans. Inf. Theory}, vol.~68, no.~7, pp. 4818--4836, Mar. 2022.

\end{thebibliography}

\end{document}